\documentstyle[11pt,leqno]{article}
 \newcommand{\beq}{\begin{equation}}                       
 \newcommand{\eeq}{\end{equation}}                         
 \newcounter{nt}[section]                                  
 \newcounter{nl}[section]                                  
 \date{}                                                   

\begin{document}

 \title{ \bf On a non linear third - order parabolic equation
 }

 \author{\sc Monica De Angelis
   \thanks{ Dipartimento di Matematica e Applicazioni, Facolt\`{a} di
             Ingegneria, via Claudio 21, 80125, Napoli.}}

 \maketitle

 \begin{abstract}

\vspace{5mm}

Aim of this paper is the qualitative analysis of the solution of a
boundary value problem for a third-order non linear parabolic equation
which describes several dissipative models. When the source term is
linear, the problem is explictly solved by means of a Fourier series
with properties of rapid convergence. In the non linear case,
appropriate estimates of this
series allow to deduce
the asymptotic
behaviour of the solution.
 \end{abstract}

  \vspace{7mm}

 \section{Introduction}
 \setcounter{equation}{0}

 \hspace{5.1mm}

We refer to the non linear equation

\vspace{3mm}
 \beq                                   \label{11}
 {\cal L}_ \varepsilon u=
F(x,t,u,u_x,u_t)
 \eeq

\vspace{3mm}\noindent
where $\ {\cal L}_ \varepsilon $ is the third - order parabolic operator:

\vspace{3mm}
 \beq                                   \label{12}
 {\cal L}_ \varepsilon =\partial_{xx}(\varepsilon
\partial_t+c^2 ) - \partial_t(\partial_t+a).
 \eeq

\vspace{3mm}
In (\ref{11})-(\ref{12})  $ \varepsilon, a, c$ are positive constants and
 $F=F(x,t,u,u_x,u_t)$  is a prefixed function.

The equation (\ref{11}) characterizes the evolution of several
dissipative models such as the
motions of viscoelastic fluids or solids, the sound propagation in
viscous gases, the heat condution at low temperature and some
Josephson effects
 in
superconductivity when dissipative causes are not neglegible (for details see, e.g, \cite{ddr}\cite{d}).

The analysis of several initial -
 boundary problems related to the equation (\ref{11}) has been
discussed in many papers (see, for example \cite{rio} and its
references).

In \cite {ddr}, the fundamental solution
 $K(x,t)$ of the operator ${\cal L}_\varepsilon$  has been determined
 explictly by means of convolutions of Bessel function, and so, when $F=f(x,t)$ is linear, the initial value problem
 has been discussed. Further, the linear strip problem has been solved in \cite {bari} by
means of a Green function G connected with
 $K(x,t)$.
 However, it seems difficult to deduce an exhaustive
asymptotic analysis by means of this solution.

In this paper, both linear and non linear cases are analyzed. The
Green function of the linear strip problem is determined by Fourier series
which is characterized by properties of rapid convergence and
allows to establish an exponential decrease of the solution.
 Moreover, the non - linear problem is reduced to
an integral equation with kernel G and, by means of suitable
properties of G, the asymptotic analysis of
the solutions is achieved.

\vspace{6mm}\setcounter{section}{1}
 \section{Some typical problems  in superconductivity}
 \setcounter{equation}{0}

 \hspace{5.1mm}

As an example of physical models related to the equation (\ref{11}), we
will consider a typical case of non linear phenomenon concerning the
Josephson effects in Superconductivity.

Let $\varphi=\varphi(x,t)$ be the phase difference of the wave functions related to the two
  superconductors, and let $\gamma$ be the normalized current bias; when
  $F=sin\varphi -\gamma$, the equation (\ref{11}) gives the perturbed Sine-
  Gordon equation \cite{bar2}:

\vspace{3mm}
 \beq                                                     \label{21}
  \varepsilon \varphi_{xxt}+\varphi_{xx}-\varphi_{tt}-a\varphi_{t}=\sin
  \varphi-\gamma.
 \eeq

\vspace {3mm}

The terms $\varepsilon \varphi_{xxt}$ and $a\varphi_t$ characterize the
 dissipative normal electron current flow along and across the junction,
 respectively,
 and they represent the
{\em perturbations} with respect to the classic Sine Gordon
 equation \cite{jos2,pag1,scott1}. When the
 surface resistence is neglegible \cite{for,nak,no}, then $\varepsilon$ is
 vanishing and a singular perturbation problem for the equation
 (\ref{11}) could be appear.
 As for the coefficient a of (\ref{21}), it depends on
 the shunt conductance \cite{sco} and generally one has $a<< 1$
 \cite{lom1,no,pag1}. However, if the resistence of the junction is so low to
 short completely the
 capacitance, the case  $a>>1$ arises \cite{bar2,tin,wa}.

If $l$ is
the normalized length of junction, a first boundary problem
 refers to the phase - boundary specifications for
$x=0$ and $x=l$  \cite{sco}. Moreover, according to the junction geometry,
other typical boundary value problems could be considered. For instance,
in an
 overlap geometry  \cite {bar2,pag1}, if  $\eta$ is the normalized measure of
the external magnetic  field applied in the plane of the junction \cite
{for},
 it has:

\vspace{3mm}
\beq                         \label {22}
\varphi_x(0,t)=\varphi _x(l,t)=\eta.
\eeq

\vspace {3mm}
Instead, if an 'open-circuit' is examined, then boundary conditions (\ref{22})
assume the homogeneous form \cite {lom1},
while, for the electrodynamics of a
Josephson junction in two spatial dimensions, other conditions on
(0,l) can be considered \cite {par}.

As an example, we will analyze the problem related to the phase boundary
specifications but the analysis can be applied to the other
problems.

\vspace{6mm}\setcounter{section}{2}
 \section{The linear strip problem}
 \setcounter{equation}{0}

\hspace{5.1mm}

Let $l , T  $ be arbitrary positive constants and let

\vspace{4mm}

$\ \ \ \ \ \ \ \ \ \ \ \  \ \ \ \ \ \ \   \Omega =\{(x,t) : 0 < x <
l, \  \ 0 < t \leq T \}$.

\vspace{4mm}\noindent
In the linear case, the strip problem is:

\vspace{3mm}
  \beq                                                     \label{31}
  \left \{
   \begin{array}{ll}
    & \partial_{xx}(\varepsilon
u _{t}+c^2 u) - \partial_t(u_{t}+au)=f(x,t),\ \  \
       (x,t)\in \Omega,\vspace{2mm}\\
   & u(x,0)=g_0(x), \  \    u_t(x,0)=g_1(x), \  \ x\in [0,l],\vspace{2mm}  \\
    & u(0,t)=0, \  \ u(l,t)=0, \  \ 0<t \leq T.
   \end{array}
  \right.
 \eeq

\vspace{3mm}

As it's well known, the problem (\ref{31}) can be solved explicitly by
means of the Fourier method, as soon
as the Green function for the operator $(\ref{31})_1$ has been determined.
For this let :

\beq                                             \label{32}
\gamma_n=\frac{n\pi}{l},\  \
\ \ b_n=c\gamma_n,\  \ \ \  h_n=\frac{1}{2}(a+\varepsilon\gamma_n^2),\  \
\  \ \omega_n=\sqrt{h_n^2-b_n^2};
\eeq

\vspace{3mm}
\noindent
and

\beq                                \label{33}
H_n(t)= \frac{1}{\omega_n}e^{-h_nt}
sinh(\omega_nt).
\eeq

\vspace{3mm}
By standard techniques, the Green function related to (\ref{31}) can be
given the form:

\vspace{3mm}
\beq                                                \label{34}
G(x,t,\xi)=\frac{2}{l}\sum_{n=1}^{\infty}
H_n(t) \  \  sin\gamma_n\xi\  \ sin\gamma_nx.
\eeq

\vspace{3mm}\noindent
whose properties will be discussed in the next section 4.

Then, if we  consider the functions:

\vspace{3mm}

\beq                                   \label{35}
u_{g_i}=\int_{0}^{l} g_i(\xi)\, G(x,\xi,t) \\\, d\xi \ \ \  \ i=0,1
\eeq

\vspace{3mm}\noindent
and

\beq                                             \label{36}
u_f=G*f=\int_{0}^{t}d\tau\ \int_{0}^{l}f(\xi,\tau)\  \ G(x,\xi,t-\tau)\
d\xi
,
\eeq

\vspace{3mm}\noindent
the formal solution of (\ref{31}) is given by:

\vspace{3mm}
\beq                                            \label{37}
u(x,t)=u_{g_1}+(\partial_t+a-\varepsilon\partial_{xx})u_{g_o}-u_f(x,t).
\eeq
\setcounter{section}{3}
 \section{Properties of the G function}
 \setcounter{equation}{0}
\hspace{4.1mm}

\vspace{5mm}
At first we will analyze the convergence and the asymptotic properties of
the series (\ref{34}) which defines $G$. We will prove that the order of the
terms of $G$ is at least  $n^{-2}$ and the function $G$
is exponentially vanishing as $t\rightarrow\infty$.
 For this we put:

\vspace{3mm}
\beq                      \label{41}
p= \frac{c^2}{\varepsilon+a(l/\pi)^2},  \ \ \  q=  \frac{a+\varepsilon
(\pi/l)^2}{2}, \ \ \ \beta\equiv min(p,q),
\eeq

\vspace{3mm}\noindent
and we will prove the following lemma:

\vspace{5mm}
{\bf Lemma 4.1}-{\em Whatever the constants} $a, \varepsilon, c^2 $
{\em may be in} $\Re^+$, {\em the function} $G(x,\xi,t)$ {\em defined in}
(\ref{34}) {\em and all its time derivatives are continuous functions in}
$\Omega$.
{\em  Moreover, everywhere in } $\Omega$, {\em  it results:}
\vspace{3mm}
\beq                                        \label{42}
|G(x,\xi,t)|\,\leq \,M  \,e^{-\beta t}, \ \ \, \ \ \ \
|\frac{\partial ^j G}{\partial t^j}|\, \leq \,N_j \,  e^{-\beta
t},
\\ \ \ \ \ \ \, \ \ j \in {\sf N}
\eeq

\vspace{3mm}

\noindent
{\em where } $M, N_j $ { \em are constants depending on } $a,\varepsilon, c^2$.

\vspace{3mm}
{\bf Proof}. When $c^2=a\varepsilon$, the operator $L_\varepsilon$ can be
reduced to the wave operator. If $c^2< a\varepsilon$, for all
$n\, \geq \,1$,
one has $h_n > b_n$.
So, if $k$ is an arbitrary constant such that
$c^2/a\varepsilon\, <\,k\,<1$, the Cauchy inequality assures
that:

\vspace{3mm}
\beq                                         \label{43}
\frac{b_n}{\sqrt{k}} \ \leq \frac{1}{2} \ (\varepsilon\gamma_n^2+a)=h_n.
\eeq

\vspace{3mm}
\noindent
Then, if $X_n =(b_n/h_n)^2$ , one has:

\vspace{3mm}
\beq                                         \label{44}
(1-k)^{\frac{1}{2}}\, \leq\,
\frac{\omega_n}{h_n}=(1-X_n)^{\frac{1}{2}}<1-X_n/2.
\eeq

\vspace{3mm}\noindent
Further, for all $n\geq1$, it results:

\vspace{3mm}

\beq                                       \label{45}
\frac{b_n^2}{2h_n}\, \\ \geq \
p,\ \ \ \ \ \ \
\ \ \ h_n>\frac{\varepsilon\pi^2}{2l^2} \, n^2 \  = (q-a/2) \ n^2
\eeq

\vspace{3mm}
\noindent
and hence, by $(\ref{44})$ and $(\ref{45})_1,$ it follows:

\vspace{3mm}

\beq                                       \label{46}
e^{-t(h_n-\omega_n)}\leq \ \ e^{-pt}.
\eeq

\vspace{3mm}
Moreover, the estimates $(\ref{44})$, $(\ref{45})_2$, (\ref{46}) imply that the terms
$H_n$ defined by (\ref{33}) verify:

\vspace{3mm}
\beq                                     \label{47}
|H_n|\ \leq \ \frac{(1-k)^{-1/2}}{q-a/2} \ \frac{e^{-pt}}{n^2},\ \
\ \ \ \forall n \ \geq 1
\eeq

\vspace{3mm}\noindent
and hence,$ (\ref{42})_1$ follows.
As for
$(\ref{42})_2$, we observe:

\vspace{3mm}
\beq                                 \label{48}
h_n-\omega_n =\frac{b_n^2}{h_n+\omega_n}\,\leq \frac{2c^2}{\varepsilon},\
\ \ \ \forall n\geq 1
\eeq

\vspace{3mm}
\noindent
and by means of standard computations one obtains $(\ref{42})_2.$

Consider now the case $c^2>a\varepsilon$, and let

\vspace{3mm}
\beq                              \label{49}
  N_{1,2} = \frac{cl}{\varepsilon\pi}[1\mp(1-a
     \varepsilon/c^2)^{1/2}].
\eeq

\vspace{3mm}
If $ k$ is a positive constant
less than one, let
 $N^*_{1,2} , N_k $ be the lowest integers such that

\vspace{3mm}
\beq                            \label{410}
N^*_1 < \ N_1, \ \ \   N^*_2 \ > N_2; \ ,\ \
     N_k \ >\frac{cl}{\varepsilon\pi\sqrt k}[1+(1-a \varepsilon
     k/c^2)^{1/2}].
\eeq

\vspace{3mm}
It suffices to analyse only the hyperbolic terms $H_n$ with
$n\geq N^*_2. $  Similarly to (\ref{44}) for  all $n \, \geq N_k,$ it results
:

\vspace{3mm}
\beq                          \label{411}
\omega_n \, \geq h_n \ (1-k)^{1/2}.
\eeq

\vspace{3mm}\noindent
So $(\ref{45})_2$,  (\ref{46}), (\ref{48}) and (\ref{411}) imply the
estimates
$(\ref{41})$ also when $c^2>a\varepsilon$.

Obviously, the lemma holds  also when $N_{1,2}$
are integers. In this case, the constants M and $N_j$ could depend on t.
$\Box$

\vspace{3mm}
As it is known, as for the x-differentiation of Fourier's series like
(\ref{34}), attention must be paid to convergence problems. For this, we
will consider x-derivatives of the operator
$(\varepsilon\partial_t+c^2)G$       instead of   $G $ and   $G_t$.

\vspace{5mm}
{\bf Lemma 4.2}.-{\em Whatever} $a , \varepsilon, c^2 $ {\em may be
, the function}   $ G(x,\xi,t)$ {\em defined in} (\ref{34}) {\em is such that:}
\vspace{3mm}

\beq                            \label{412}
|\varepsilon\, G_t+c^2G| \\ \, \leq A_o \, \,  e^{-\beta t}
,
\eeq

\beq                            \label{413}
|\partial_{xx}(\varepsilon\, G_t\, +c^2\, G )|\\ \,\leq A_1 \, \,
\, e^{-\beta t},
\eeq
\vspace{3mm}

\noindent
{\em where $A_0$ and $ A_1$  are constants depending  on $a, \varepsilon ,
c^2 $.}

\vspace{3mm}
{\bf Proof}.
As for the
hyperbolic terms in $H_n, $ we consider $X_n=(\frac{b_n}{h_n})^2<1$  and put:

\vspace{3mm}
\beq                                      \label{414}
\varphi_n=h_n(-1+\sqrt{1-X_n}).
\eeq

\vspace{3mm}\noindent
It results:

\vspace{3mm}
\beq                                    \label{415}
\ \ \ \ \ \ \ \ \ \varepsilon
\dot H_n+c^2H_n=\frac{1}{2\omega_n}e^{-h_nt}\{(c^2+\varepsilon\varphi_n)
e^{\omega_nt}-[c^2-\varepsilon(h_n+\omega_n)]e^{-\omega_nt}\},
\eeq

\vspace{3mm}\noindent
where Taylor's formula assures that:

\vspace{3mm}
\beq                         \label{416}
c^2+\varepsilon \ \varphi_n
=\frac{c^2a}{a+\varepsilon\gamma_n^2}-\varepsilon
h_nX_n^2-\frac{3}{16}\int_0^{X_n}\frac{(X_n-y)^2}{(1-y)^{5/2}} \,\  dy.
\eeq

\vspace{3mm}\noindent
Further, it is possible to prove that, for all n, it is:

\vspace{3mm}
\beq                         \label{417}
X_n < \, \frac{\alpha^2}{n^2}\ \  \, \, \ \ \ \ \ (\alpha =
2cl/\varepsilon\pi)
\eeq

\vspace{3mm}\noindent
while, for every $h>0$ and  $n\geq\alpha(1+h)$, one has:

\beq                       \label{418}
\int_0^{X_n}\frac{(X_n-y)^2}{(1-y)^{5/2}} \,\  dy\, \leq \frac{2}{3} \ X_n^2 \,
\{(h+1)^3/[h(h+2)]^{3/2}-1\}.
\eeq

\vspace {3mm}\noindent
So, estimates (\ref{417}) and (\ref{418}) allow to obtain:

\vspace{3mm}
\beq                         \label{419}
|c^2+\varepsilon \ \varphi_n | \ \, \leq \frac{1}{n^2} \
(5\alpha^2/4+\alpha_1/8),
\eeq

\vspace{3mm}\noindent
where $\alpha_1=\alpha^4[(h+1)^3/(h^2+2h)^{3/2}-1]$.

\noindent
The estimate of Lemma 4.1 together with (\ref{419}) show that the terms of the serie related to the operator (\ref{412}) have order at
least  of $n^{-4}$. So, it can be differentiated term by terms with respect
to x and the estimate (\ref{413}) can be deduced.
$\Box$

\vspace{5mm}
As solution of the equation
${\cal L}_\varepsilon v
 = 0$  we will mean a continuous function
$v(x,t)$ which has continuous the derivatives $v_t, v_{tt}, $ $\partial_{xx}(\varepsilon
v_t+c^2v) $  and these derivatives verify the equation.

\noindent
So,
 we are able to prove the following theorem:

\vspace {5mm}
{\bf Theorem 4.1}- {\em The
function  $G(x,t)$ defined in} (\ref{34}){\em is a
solution of the equation }

\vspace{3mm}
\beq                      \label{420}
{\cal L}_\varepsilon G \, =\partial_{xx}(\varepsilon
G _{t}+c^2 G) - \partial_t(G_{t}+aG)=0.
\eeq

\vspace{3mm}
{\bf Proof}. The uniformly convergence proved in Lemma 4.1 and Lemma 4.2
allow to deduce that:

\vspace{3mm}
\beq                      \label{421}
(\partial_t+a)G_t=
\frac{2}{l}\\\sum_{n=1}^{\infty}\{[-b_n^2+(2h_n-a)h_n]H_n+(a-2h_n)
e^{-h_nt}\cosh\omega_nt\}\sin\gamma_n\xi
\sin
\gamma_nx,
\eeq

\vspace{3mm}\noindent
and

\beq                           \label{422}
 \ \ \ \ \ \ \ \ \ \partial_{xx}(\varepsilon\partial_t+c^2)G=\frac{2}{l}\\\sum_{n=1}^{\infty}[(\varepsilon\gamma^2_nh_n-b^2_n)H_n-\varepsilon\gamma^2_n
e^{-h_nt} \cosh\omega_nt]\sin\gamma_n\xi
\sin
\gamma_nx,
\eeq

\vspace{3mm}\noindent
and hence, (\ref{420}) follows.$\Box$

\vspace{6mm}\setcounter{section}{4}
 \section{Properties of the convolution}
 \setcounter{equation}{0}

\hspace{5.1mm}

To obtain the explicit solution of the linear strip problem, the
properties of the convolution of function G with the data must be
analyzed.

\noindent
For this, let $g(x) $ be a continuous function on $(0,l)$ and let consider:

\vspace{3mm}
\beq                             \label{51}
u_g(x,t)\ \ =
\int_{0}^{l} g(\xi) \ \ G(x,\xi,t)\ \ d\xi.
\eeq

\beq                             \label{52}
u_g^*(x,t)\ \ =
(\partial_t+a-\varepsilon\partial_{xx})u_g(x,t).
\eeq

\vspace{3mm}\noindent
So the following lemma holds:

\vspace{5mm}
{\bf Lemma 5.1}- {\em If} $g(x)$ {\em is a} $C^1(0,l),$
{\em then
the function} $u_g$ {\em defined by} (\ref{51}) {\em is a solution}
(n.4) {\em  of the
equation} ${\cal L}_\varepsilon =0$ {\em such that}:

\vspace{3mm}
\beq                            \label{53}
\lim_{t \rightarrow 0}u_g(x,t)=0 \ \ \ \ \ \lim_{t \rightarrow 0} \partial_t
u_g(x,t)=g(x),
\eeq

\vspace{3mm}\noindent
{\em uniformly for all }$x\in (0,l)$.

\vspace{3mm}
{\bf Proof.} Lemma (\ref{41})-(\ref{42}) and continuity of g assure
that function
(\ref{51}) and the partial derivatives required by the solution converge absolutely for all
$(x,t)\in \Omega$. Hence, by theorem 4.1, one deduces:

\vspace{3mm}
\beq                             \label{54}
\ \ \ \ \ \ \\\ \partial_{xx}(\varepsilon\partial_t+c^2)u_g=\int_{0}^{l}
g(\xi) \ \partial_{xx}(\varepsilon\partial_t+c^2)G(x,\xi,t) \  d\xi=
\partial_t(\partial_t+a)u_g.
\eeq

\vspace{3mm}\noindent
Besides, the hypoteses on the function
g and (\ref{42}), imply $(\ref{53})_1$.

Moreover, one has:

\vspace{3mm}
\beq                            \label{55}
G_t=-\frac{2}{\pi} \ \ \frac{\partial}{\partial\xi} \ \ \sum_{n=1}^{\infty}
\ \ \dot H_n(t) \frac{\cos \gamma_n\xi}{n}\sin
\gamma_n x,
\eeq

\vspace{3mm}\noindent
and hence:

\vspace{3mm}
\beq                            \label{56}
\partial_tu_g= -\frac{2}{\pi}
\sum_{n=1}^{\infty}
\ \ \frac{\dot H_n(t)}{n} \ \ [g(\xi)\cos \gamma_n\xi \
]^l_0 \ \ \sin\gamma_n x \ + \eeq
\hspace{4cm}\[+\frac{2}{\pi} \int_{0}^{l}\sum_{n=1}^{\infty}
\ \ \dot H_n(t)  \,g^{'}(\xi) \, \frac{\cos \gamma_n\xi}{n} \,\sin
\gamma_n x \ d\xi.\]

\vspace{3mm}
So, if $\eta(x)$ is the Heaviside function, from (\ref{56}), one obtains:

\vspace{3mm}
\beq                            \label{57}
\ \ \ \ \ \ \ \ \ \lim_{t \rightarrow 0}\partial_tu_g=
\frac{x}{l}[g(l)-g(0)]+g(0)- \int_{0}^{l} g^{'}(\xi) [\eta(\xi-x)+\frac{x}{l}-1]
d\xi  =g(x).
\Box
\eeq

\vspace{5mm}
{\bf Lemma 5.2}- {\em If} $g(x) \in C^3(0,l)$ {\em with} $g^{(i)}(0)=g^{(i)}(l)=0
  (i=1,2,3)$,
{\em
then the function} $u_g^*$ {\em defined in }(\ref{52}) {\em is a solution
(n.4) of the
equation} ${\cal L}_\varepsilon =0$ {\em such that}:

\vspace{3mm}
\beq                            \label{58}
\lim_{t \rightarrow 0}u_g^*(x,t)=g \ \ \ \ \ \lim_{t \rightarrow 0} \partial_t
u_g^*(x,t)=0
\eeq

\vspace{3mm}\noindent
{\em uniformly for all }$x\in (0,l)$.

\vspace{3mm}
{\bf Proof.} The hypoteses on $g(x)$
allow to have

\vspace {3mm}
\beq                               \label{59}
\partial_{xx}u_g(x,t) =
\int_{0}^{l} g^{''}(\xi) \ \ G(x,\xi,t)\ \ d\xi =  u_{g^{''}}(x,t).
\eeq

\vspace{3mm}\noindent
So, by Lemma 5.1 and expression (\ref{59}), one obtains
:

\vspace {3mm}
\beq                               \label{510}
\ \ \ \partial_t(\partial_t+a)u_g^* =
\partial_t(\partial_t+a)[(\partial_t+a)u_g-\varepsilon u_{g^{''}}]= \partial_{xx}(\varepsilon\partial_t+c^2)u_g^*.
\eeq

\vspace{3mm}\noindent
Moreover,
it results:

\vspace{3mm}
\beq                                    \label{511}
\partial_t u_g^* = \partial^2_t
u_g+a\partial_tu_g-\varepsilon\partial_tu_{g^{''}} = c^2u_{g^{''}},
\eeq

\vspace{3mm}\noindent
and hence,
$(\ref{53})_1$  implies $(\ref{58})_2. $

\noindent
Finally, owing to (\ref{53})-(\ref{59}), one has:

\vspace{3mm}
\beq                            \label{512}
\lim_{t \rightarrow 0}u_g^*=\lim_{t \rightarrow 0}
[(\partial_t+a)u_g - \varepsilon u_{g^{''}}] = g(x).\Box
\eeq

\vspace{6mm}\setcounter{section}{5}
 \section{Explicit solution of the strip linear problem}
 \setcounter{equation}{0}

\hspace{5.1mm}

\vspace {3mm}
Let us consider the homogeneous case .
>From Lemma 5.1 and Lemma 5.2 the following result is obtained:

\vspace{5mm}
{\bf Theorem 6.1}- {\em If the initial data}  $g_1(x),$ {\em and }$   g_0(x)$ {\em verify
the hypotheses of Lemma 5.1 and Lemma 5.2,} {\em then the function}:

\vspace{3mm}
\beq                        \label{61}
u(x,t)=u_{g_1}+(\partial_t+a-\varepsilon\partial_{xx})u_{g_0}
\eeq

\vspace{3mm}\noindent
{\em represents a solution} (n.4) {\em of the homogeneous strip problem:}

\vspace{3mm}
  \beq                                                     \label{62}
  \left \{
   \begin{array}{ll}
    & \partial_{xx}(\varepsilon
u _{t}+c^2 u) - \partial_t(u_{t}+au)=0,\ \  \
       (x,t)\in \Omega,\vspace{2mm}\\
   & u(x,0)=g_0(x), \  \    u_t(x,0)=g_1(x), \  \ x\in [0,l],\vspace{2mm}  \\
    & u(0,t)=0, \  \ u(l,t)=0, \  \ t \in (0,T].
   \end{array}
  \right.
 \eeq

\vspace{3mm}
As for the non-homogeneous problem, let:

\vspace{3mm}
\beq                             \label{63}
u_f(x,t)\ \ = \int_{0}^{t} d\tau
\int_{0}^{l} f(\xi,\tau) \ \ G(x,\xi,t-\tau)\ \ d\xi,
\eeq

\vspace{3mm}\noindent
the following theorem holds:

\vspace{5mm}
{\bf Theorem 6.2}- {\em If the function }  $f(x,t)$ {\em
is a continuous function
in $\Omega$ with continuous derivative with respect to x, }
 {\em then the function} $u_f$
{\em represents a solution} (n.4) {\em of the non-homogeneous strip problem:}

\vspace{3mm}
  \beq                                                     \label{64}
  \left \{
   \begin{array}{ll}
    & \partial_{xx}(\varepsilon
\partial_t+c^2) u - \partial_t(\partial_t+a)u=-f(x,t),\ \  \
       (x,t)\in \Omega,\vspace{2mm}\\
   & u(x,0)=0, \  \    \partial_tu(x,0)=0, \  \ x\in [0,l],\vspace{2mm}  \\
    & u(0,t)=0, \  \ u(l,t)=0, \  \ t \in (0,T].
   \end{array}
  \right.
 \eeq

\vspace{3mm}
{\bf Proof.}
By (\ref{34}) one has:

\vspace{3mm}
\beq                            \label{65}
\ \ \ \lim_{\tau \rightarrow t}u_f(x,t)=0,
\eeq

\vspace{3mm}\noindent
and hence:

\vspace{3mm}
\beq                             \label{66}
\partial_tu_f(x,t)\ \ = \int_{0}^{t} d\tau
\int_{0}^{l} f(\xi,\tau) \ \ G_t(x,\xi,t-\tau)\ \ d\xi.
\eeq

\vspace{3mm}
Moreover, as in (\ref{57}), it is possible to prove that:

\vspace{3mm}
\beq                            \label{67}
\lim_{\tau \rightarrow t}
\int_{0}^{l} f(\xi,\tau) \ \ G_t(x,\xi,t-\tau)\ \ d\xi = f(x,t),
\eeq

\vspace{3mm}\noindent
and one obtains:

\vspace{3mm}
\beq                             \label{68}
\partial^2_tu_f\ \ = f(x,t)+
\ \int_{0}^{t} d\tau
\int_{0}^{l} f(\xi,\tau) \ \ G_{tt}(x,\xi,t-\tau)\ \ d\xi.
\eeq

\vspace{3mm}
So that, Lemma 4.2, Theorem 4.1 and (\ref{64})-(\ref{66}) assure that:

\vspace{3mm}
\beq                         \label{69}
\ \ \ \ \ \ \ \ \partial_{xx}(\varepsilon\partial_t+c^2)u_f=\int_{0}^{t}d\tau
\int_{0}^{l} f(\xi,\tau) \partial_{xx}(\varepsilon G_{t}+c^2 G)
d\xi=\partial_t(\partial_t+a)u_f-f(x,t).
\eeq

\vspace{3mm}\noindent
Finally,
owing to (\ref{63})-(\ref{66}) and Lemma 4.1, if $B_i$ (i=1,2) are two
positive constants, one has:

\vspace{3mm}
\beq                              \label{610}
|u_f|\leq   B_1 (1-e^{-\beta t}); \ \ \ \\ \  |\partial_tu_f|\leq
B_2 (1-e^{-\beta t}), \eeq

\vspace{3mm}\noindent
from which $(\ref{64})_2$ follow.$\Box$

\vspace{3mm}
Moreover, the uniqueness is a consequence of the energy-method (see, for
example, \cite {dr}) and we
have:

\vspace{5mm}
{\bf Theorem 6.3}- {\em When the source term }  $f(x,t)$ {\em satisfies
theorem 6.2, and the initial data} $(g_0,g_1)$ {\em satisfy theorem 6.1,}
 {\em then the function}

\beq                                \label{611}
u(x,t)=u_{g_1}+(\partial_t+a-\varepsilon\partial_{xx})u_{g_0}-u_f
\eeq
{\em is the unique solution} (n.4) {\em of the linear non-homogeneous
strip problem} (\ref{31})$\Box$

\vspace{5.1mm}
\setcounter{section}{6}
 \section{Asymptotic properties}
 \setcounter{equation}{0}

\hspace{5.1mm}

At first we will consider the linear problem (\ref{31}). Obviously, the
asymptotic properties of the solution depend on the
behaviour of the source term $f=
f(x,t)$. For instance, one has:

\vspace{5mm}
{\bf Theorem 7.1}- {\em When the source term } $f(x,t)$ {\em satisfies the
condition}:

\vspace{3mm}
\beq                   \label{71}
|\, f(x,t)\,| \, \leq\, h \, \frac{1}{(k+t)^{1+\alpha}}\\\  \ \\ \ \ \\ \
\ \ \ \ \ \ \  \ \\ \ \ \  \ \ \
(h, k, \alpha = const > 0),
\eeq

\vspace{3mm}\noindent
{\em then the solution of} (\ref{31}) {\em is vanishing
as} t $\rightarrow  \infty$,
{\em at least as} $t^{-\alpha}$.

{\em Moreover, when}

\vspace{3mm}
\beq                   \label{72}
|\, f(x,t)\,| \, \leq\, C \, e^{-\delta t}  \ \\ \ \ \\ \\ \ \ \ \ \ \ \ \
\
(C, \delta = const > 0),
\eeq

\vspace{3mm}\noindent
{\em one has}:

\vspace{3mm}
\beq                  \label{73}
|\, u(x,t)\,| \, \leq\, k \, e^{-\delta^* t}  \ \\ \ \ \\ \\   \  \ \
\delta^* = min\{\beta, \delta\}, \ \ k=const.
\eeq

\vspace{3mm}
{\bf Proof.}
Referring to the explicit solution  (\ref{611}) of the problem
(\ref{31}), we observe that, owing to the Lemma 4.1, the terms $u_{g_0}$ and  $u_{g_1}$  defined by
(\ref{51}) are exponentially vanishing as  t  $\rightarrow  \infty $
. Moreover, the convolution $u_f$ given by (\ref{36}) can be
estimated by (\ref{42}) too and the hypotheses on $f(x,t) $ imply the
statement.$\Box$

\vspace{6mm}
Consider now the non linear problem:

\vspace{3mm}
  \beq                                                     \label{74}
  \left \{
   \begin{array}{ll}
    & \varepsilon u_{xxt}+c^2u_{xx}-u_{tt}-au_{t}=F(x,t,u), \  \
       \vspace{2mm}\\
   & u(x,0)=g_0, \  \    u_t(x,0)=g_1, \  \ \vspace{2mm}  \\
    & u(0,t)=0, \  \ u(l,t)=0. \  \
   \end{array}
  \right.
 \eeq

\vspace{3mm}

When one applies the results of section 6 (see  \ref{63}), one obtains  the
following integral equation:

\vspace{3mm}
\beq                                          \label{75}
\ \ \ \ \  \\\\\ \ \ u(x,t)=\,
\int_{0}^{l} g_1(\xi) G(x,\xi,t) d\xi
+
(\partial_t+a-\varepsilon\partial_{xx})\int_{0}^{l} g_0(\xi)
G(x,\xi,t) d\xi +
\eeq
\hspace*{2cm}\[\int_0^ t d\tau\, \int_0^l\, G(x,\xi,t-\tau)\,
F(\xi,\tau,u(\xi,\tau))d\xi,\]

\vspace{3mm}\noindent
to which the previous estimates can be applied.

For instance, as for the perturbed Sine Gordon equation (\ref{21}),
one has:

\vspace{3mm}
\beq                                    \label {76}
|F(x,t,u)|=|sin u  +\gamma| \ \leq \gamma_1,  \ \ \ \ \  \gamma_1
>0
\eeq

\vspace{3mm}\noindent
and so the Lemma 4.1 implies that the solution of the related strip
problem is bounded for all t.

\vspace{3mm}

Another example of applications could be given by exponentially
vanishing non linear source F(x,t,u(x,t)) such that (see \cite{ce}):

\vspace{3mm}
\beq                                    \label {77}
|F(x,t,u(x,t))| \leq \ \ const.  \ \ e^{-\mu t} \ \ \ \ \\\\\\\\\\\\ (\mu
>0)
\eeq

\vspace{3mm}
As consequence of (\ref{75}) and (\ref{42}), the solution of
(\ref{74}) will vanish exponentially, too.

\vspace {10mm}

 {\bf Acknowledgement.} \ This research was supported by Italian Ministry of
 University and Scientific Research.

\begin{thebibliography}{99}


 \bibitem{bar2} A. Barone, G. Patern\`o, {\it Physics and applications of the
 Josephson effects}, Wiley (1982).

\bibitem{ce} T.K. Caughey, J. Ellison ,{\it Existence, uniqueness and
stability of solutions of a class of non linear partial differential
equation}, J.Math An. Appl. {\bf 51} (1975).





\bibitem{lom1} P.L. Christiansen, J.C.Eilbeck P.S.Lomdahl,
 A.C.Scott, H.Soerensen {\it Multiple frequency generation by bunched solitons in
Josephson tunnel junctions }, Phy Rew B 24,12  (1981).


\bibitem{for} P. L. Christiansen, M. G. Forest, S. Pagano, R. D. Parmentier,
   M. P. Soerensen, S. P. Sheu, {\it Numerical evidence for global bifurcations
   leading to switching phenomena in long Josephson junctions}, Wave Motion,
   {\bf 12},  (1990).





\bibitem{sco} F. Y. Chu, A. C. Scott, S. A. Reible, {\it Magnetic-flux
  propagation on a Josephson transmission}, J. Appl. Phys. {\bf 47}, (7)
 (1976).


\bibitem{ddr} B. D'Acunto, M. De Angelis, P. Renno, {\it Fundamental
solution of a dissipative operator}, Rend. Acc. Sc. Fis. Mat. (1997)

\bibitem{d} M. De Angelis {\it On he perturbed Sine Gordon Equation} Workshop " Modelli fluidodinamici per le applicazioni" (1996)
\bibitem{bari} B. D'Acunto, M. De Angelis, P. Renno, {\it Estimates for
the perturbed Sine Gordon equation
}, Rend. Cir Mat Pal. serieII Supp 57(1998)



\bibitem{dr} B. D'Acunto, P. Renno, {\it On the operator $\varepsilon
\partial_{xxt}+c^2\partial_{xx}-\partial_{tt}$ in general domains
,} Atti Sem Mat Fis Univ Modena , {\bf XLVII}, 1999.

\bibitem{rio} N. Flavin, S. Rionero {\it Qualitative Estimates for Partial
Differential Equations}, CRC Press, (1995).



 \bibitem{jos2} B.D. Josephson, {\it Supercurrents through barriers},
Advances in Phisics vol.14 (1965).

 \bibitem{nak} K. Nakajima, Y. Onodera, T. Nakamura, R. Stato, {\it Numerical
analysis of vortex motion on Josephson Structures}, J. Appl. Phys.,
{\bf 45} (9) (1974).

 \bibitem{no}  K.Nakajima, Y.Onodera, Y.Ogama, {\it Logic design of
Josephson network}, J. Appl. Phys.,{\bf 47}  (4) (1976)



 \bibitem{pag1} S. Pagano, {\it Licentiate Thesis DCAMM}, Reports 42, Teach
    Univ. Denmark Lyngby Denmark, (1987), (unpublished).

 \bibitem{par} R. D. Parmentier, {\it Solitons and long Josephson junctions},
    The New Superconducting Electronics, Kluver Academic Publisher,
   (1993).





\bibitem{scott1} A.Scott,{\it Active and non linear wave propagation in
electronics}, Wiley Interscience, (1969).



 \bibitem{tin} M. Tinklar, {\it Introduction to superconductivity} McGraw-Hill
    (1996).

\bibitem{wa} J.R. Waldram, A.B. Pippard,F.R.S., J.Clarke, {\it theory of
the current-voltage characteristics of S N S junctions and other
superconducting weak links}, Phil.Trans Roy. Soc. Lon {\bf A 268} (1970)




 \end{thebibliography}
\end{document}